\documentclass{aa}
\usepackage{graphicx}
\usepackage{txfonts}
\usepackage{natbib}
\bibpunct{(}{)}{;}{a}{}{'}
\begin{document}
\authorrunning{O.Pani\'c et al.}
\titlerunning{Gas and dust mass in the disk around \\
the Herbig Ae star HD169142}
\title{Gas and dust mass in the disk around the Herbig Ae star HD169142}
\author{O. Pani\'c\inst{1} \and M.R. Hogerheijde\inst{1} \and D. Wilner\inst{2} \and C. Qi\inst{2}}
\institute{Leiden Observatory, Leiden University, P.O.Box 9513, 2300 RA, Leiden, The Netherlands
           \and
            Harvard-Smithsonian Center for Astrophysics, 60 Garden Street, Cambridge, MA 02138, USA 
}
\date{Received  ; Accepted }

\abstract
{Spatially resolved observations of circumstellar disks at millimeter wavelengths allow detailed comparison with theoretical models for the radial and vertical distribution of the material.}
{We investigate the physical structure of the gas component of the disk around the pre-main-sequence star \object{HD169142} and test the disk model derived from the spectral energy distribution.}
{The $^{13}$CO and C$^{18}$O $J$=2--1 line emission is observed from the disk with $1\farcs4$ resolution using the Submillimeter Array. 
We adopt the disk physical structure derived from a model which fits the spectral energy distribution of HD169142. We obtain the full three-dimensional information on the CO emission with the aid of a molecular excitation and radiative transfer code. This information is used for the analysis of our observations and previous $^{12}$CO $J$=2--1 and 1.3~mm continuum data.}
{The spatially resolved $^{13}$CO and C$^{18}$O emission shows a Keplerian velocity pattern 
The disk is seen at an inclination close to 13$^{\circ}$ from face-on.
We conclude that the regions traced by different CO isotopologues 
are distinct in terms of their vertical location within the disk, their temperature and their column densities. With the given disk structure, we find that freeze-out is not efficient enough to remove a significant amount of CO from gas phase. Both observed lines match the model prediction both in flux and in the spatial structure of the emission. Therefore we use our data to derive the $^{13}$CO and C$^{18}$O mass and consequently the $^{12}$CO mass using standard isotopic ratios.} 
{We constrain the total disk gas mass to (0.6-3.0)$\times$10$^{-2}$~M$_{\odot}$.
Adopting a maximum dust opacity of 2~cm$^2$g$^{-1}_{\rm dust}$ we derive a minimum dust mass of 2.16$\times$10$^{-4}$~M$_{\odot}$ from the fit to the 1.3~mm data. 
Comparison of the derived gas and dust mass shows that the gas to dust mass ratio of 100 is only possible under the assumption of a 
dust opacity of 2~cm$^{2}$/g$^{-1}$ and 
$^{12}$CO abundance of 10$^{-4}$ with respect to H$_2$. However, our data are also compatible with a gas to dust ratio of 25, with a dust opacity of 1~cm$^{2}$/g$^{-1}$ and $^{12}$CO abundance of 2$\times$10$^{-4}$.}

{}

\titlerunning{Gas and dust mass in the disk around the Herbig Ae star HD169142}
\authorrunning{O. Pani\'c et al.}
\maketitle

\begin{figure*}
\centering
\includegraphics[angle=-90,width=17cm]{mom.ps}
\centering
\caption
	{Integrated intensity (contours) and first moment maps (colour scale) of $^{12}$CO J$=$2--1 (left panel, from \citet{raman}), $^{13}$CO J$=$2--1 (middle panel) and C$^{18}$O $J$=2--1 line (right panel). Contours are 1, 2, 3,...$\times$200~mJy~beam$^{-1}$~km~s$^{-1}$ for $^{12}$CO and $^{13}$CO, and 1, 2, 3,...$\times$100~mJy~beam$^{-1}$~km~s$^{-1}$ for C$^{18}$O. The integrated intensity and first moment maps are obtained over a velocity range of 5.6-8.4~km~s$^{-1}$. The data were clipped at 0.7, 0.5, and 0.35~Jy~beam$^{-1}$ for $^{12}$CO, $^{13}$CO and C$^{18}$O, respectively.}
 \label{mom} 
\end{figure*}

\section{Introduction}
\begin{figure}[!htp]
\includegraphics[angle=0,width=8cm]{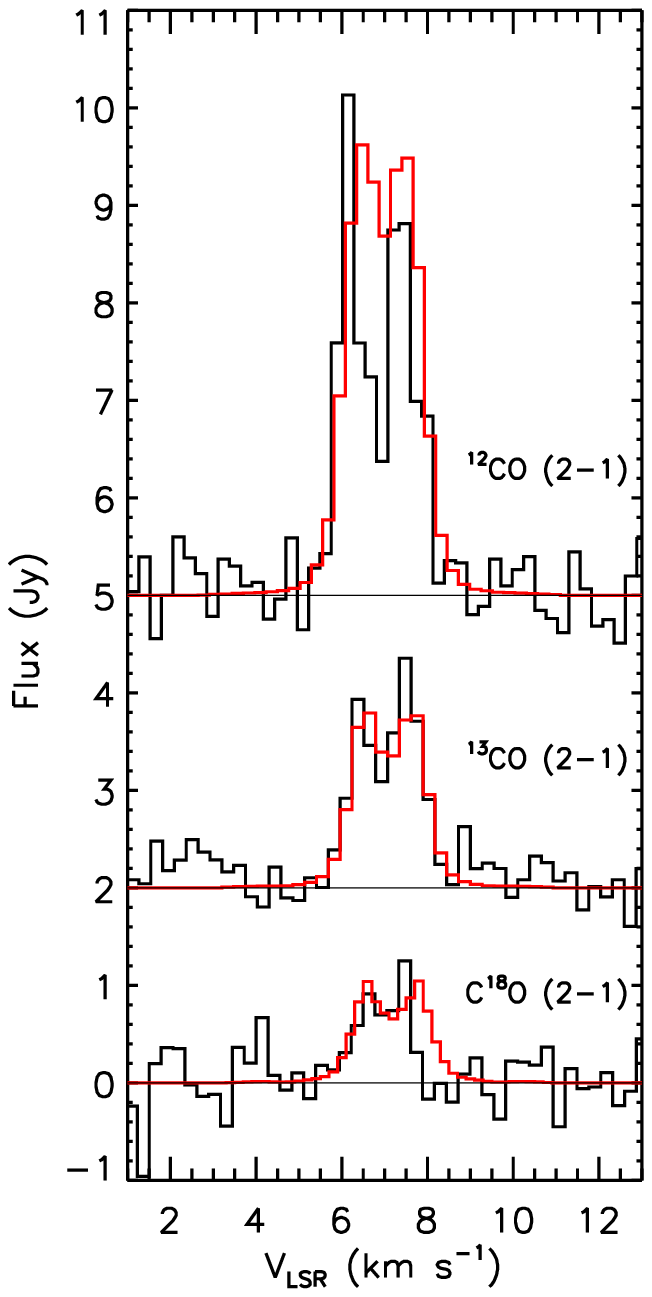}
\caption
      {Spectra of  C$^{18}$O J$=$2--1 (bottom), $^{13}$CO J$=$2--1 (middle) and $^{12}$CO $J$=2--1 line (top, from \citet{raman}) summed over the central 4$\arcsec\times$4$\arcsec$ toward HD169142. The $^{13}$CO and $^{12}$CO fluxes are shifted vertically by 2 and 5~Jy respectively. 
Black lines show the observed spectra and the red lines show the model fit found in Section 4.3.1 (for $^{12}$CO the model from \citet{raman} was used).}
\label{spec}
\end{figure}

Although the presence of molecular gas in disks around intermediate mass pre-main-sequence (Herbig Ae) stars was reported a decade ago \citep{mannings}, the research in this field has focused more on their less massive counterparts (T Tauri stars). The disks around T Tauri stars have masses ranging from 0.001 to 0.1~M$_{\odot}$ \citep{beckwith}, usually derived from millimeter continuum fluxes assuming a gas to dust mass ratio of 100, as in molecular clouds. Their outer radii are constrained by molecular line observations and are typically a few hundred AU \citep[][and references therein]{simon,thi}. Due to the low luminosity of the central star (0.5 to 1~L$_{\odot}$), these disks are relatively cold (less than 20~K beyond 100~AU from the star) causing a significant depletion of the CO in the midplane of the outer disk. On the other hand, the Herbig Ae stars are about ten times more luminous than T Tauri stars, and consequently their disks are warmer. This allows the easiest to detect and the most commonly used gas tracer, carbon monoxide, to remain in the gas phase even in the disk midplane. Observations of CO and its isotopologues toward Herbig Ae stars are therefore expected to be more powerful probes of the full disk structure. Only a few Herbig Ae disks have been studied thoroughly via spatially resolved observations of molecular line emission that includes the optically thin CO isotopologues:  AB Aur \citep{pietu2}, MWC480 \citep{pietu1}, HD163296 \citep{isella}.

\section{HD169142}

The object of our study, HD169142, is a 2.0~M$_{\odot}$ Herbig Ae star of spectral type A5Ve surrounded by a gas-rich circumstellar disk located at 145~pc \citep{sylvester}. 
With an age of 6$^{+6}_{-3}$~Myr \citep{grady} and its spectral energy distribution marked by infrared excess and the lack of silicate features \citep{dentHD}, HD169142 is an example of an advanced pre-main-sequence evolutionary stage. Unlike most of the Herbig Ae/Be stars, it shows no evidence of proximity to a cloud or extended molecular gas \citep{meeus}. Observations of molecular gas in this disk are therefore easier to interpret. However, HD169142 is not completely isolated from other young stars: \citet{grady} find three coeval pre-main-sequence stars within a projected separation of 1160~AU. The closest companion is located at 9$\farcs$3 separation and may form a binary system with HD169142. Near-infrared polarisation images show that the dust in the disk extends to at least 217~AU \citep{kuhn}. More recent submillimeter observations \citep{dentHAe,raman} show bright and narrow CO lines. \citet{raman} spatially resolve the disk and find a fit to the CO $J$=2--1 line and 1.3~mm continuum observations by adopting a flared accretion disk model with a 235~AU radius and a 13$^{\circ}$ inclination from face on. Observations at optical, IR, and (sub)millimeter wavelengths allowed  modelling of the disk's spectral energy distribution (SED) \citep{malfait,dominik,dentHD}. \citet{malfait} fitted the near-infrared and far-infrared excess of HD169142 by two disk components: an inner disk extending from 0.5~AU to 1~AU with a density exponent of 2.0 and the outer disk from 28~AU with a flatter density distribution. \citet{dominik} adopt a low inclination of 8$^{\circ}$, outer radius of 100~AU and surface density exponent $p$=2 to fit the SED, and therefore derive a disk mass of 0.1~M$_{\odot}$. A more detailed SED modeling is done by \citet{dentHD} where both the SED and resolved 7~mm continuum emission were fitted using an accretion disk model \citep{dalessio} corresponding to a 10~Myr old A2 spectral type star. They adopt an inclination of 30$^{\circ}$ and outer radius of 300~AU, and derive a disk mass of 4$\times$10$^{-2}$~M$_{\odot}$.  \citet{grady} fit the SED and NICMOS image at 1.1~$\mu$m with a model consisting of two distinct disk components - the inner disk from 0.15 to 5~AU radius and the outer disk extending from 44 to 230~AU. It is important to stress that all above mass estimates of the disk around HD169142 are based solely on the observed dust emission, and not gas.

This paper presents resolved interferometric observations of the $^{13}$CO and C$^{18}$O $J$=2--1 lines from HD169142. The observations and results are shown in Sections~3 and 4. Section~4 introduces the disk model we adopt \citep{dalessio,dentHD,raman} and our fit to the 1.3~mm data providing a measure of the minimum disk dust mass. We fit the $^{13}$CO and C$^{18}$O emission deriving the corresponding $^{13}$CO and C$^{18}$O mass in the disk, and place constraints on the total gas mass of the disk. We discuss the implications of our findings for the gas to dust ratio in the disk and the possible presence of substructure in the disk midplane.
Section~5 summarizes our findings.

\section{Observations and results}
The observations of HD169142 were carried out with the Submillimeter Array\footnote{The Submillimeter Array is a joint project between the Smithsonian Astrophysical Observatory and the Academia Sinica Institute of Astronomy and
Astrophysics and is funded by the Smithsonian Institution and the Academia Sinica.} (SMA) on 2005 April 19, simultaneous with the observations of $^{12}$CO J=2--1 line presented in \citet{raman}. A more detailed description of the observations and of the calibration procedure is given there. 
The correlator provided $2~$GHz of bandwidth in each sideband and was 
configured to include the $^{13}$CO J=2--1 line at 220.3986765~GHz and the C$^{18}$O
 J=2--1 line at 219.5603568~GHz in the lower sideband in a 104~MHz wide spectral band 
with channel spacing of 0.2~MHz ($\sim$0.26~km~s$^{-1}$). 

The data reduction and image analysis were done with the Miriad data reduction tools \citep{sault}. The (u,v) data were Fourier transformed using natural weighting. The resulting synthesized beam size is $1\farcs4\times1\farcs0$ (PA=26$^{\circ}$). The rms of the line images is 180~mJy~beam$^{-1}$ per channel or 4.6~K (4.8~K for $^{12}$CO). 

\begin{figure*}[!htp]
\centering
\includegraphics[width=17cm]{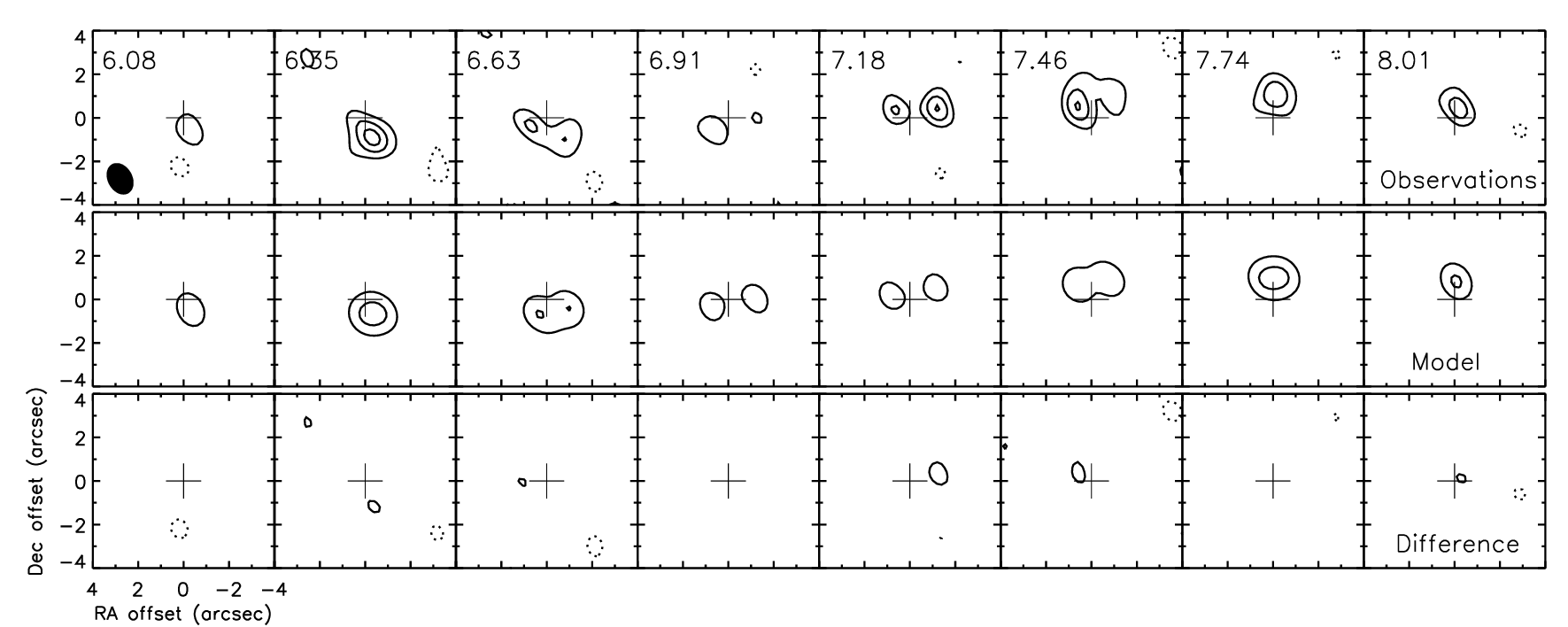}
\caption
      {-- {\it{Top panel}:} Channel maps of the observed $^{13}$CO $J$=2--1 emission at a range of velocities over which the line is detected. The lower left corner of this panel shows the size and position angle of the synthesized beam. -- {\it{Middle panel}:} $^{13}$CO $J$=2--1 chanel maps from the best-fit disk model from Section 4.3.1. --  {\it{Lower panel}:} Channel maps showing the residual emission after subtraction of the best-fit model from the data. The contour levels are -1,1,2,3,4$\times$360~mJy~beam$^{-1}$ (2 sigma) in all panels.}
 \label{13maps} 
\end{figure*}
\begin{figure*}
\centering
\includegraphics[width=17cm]{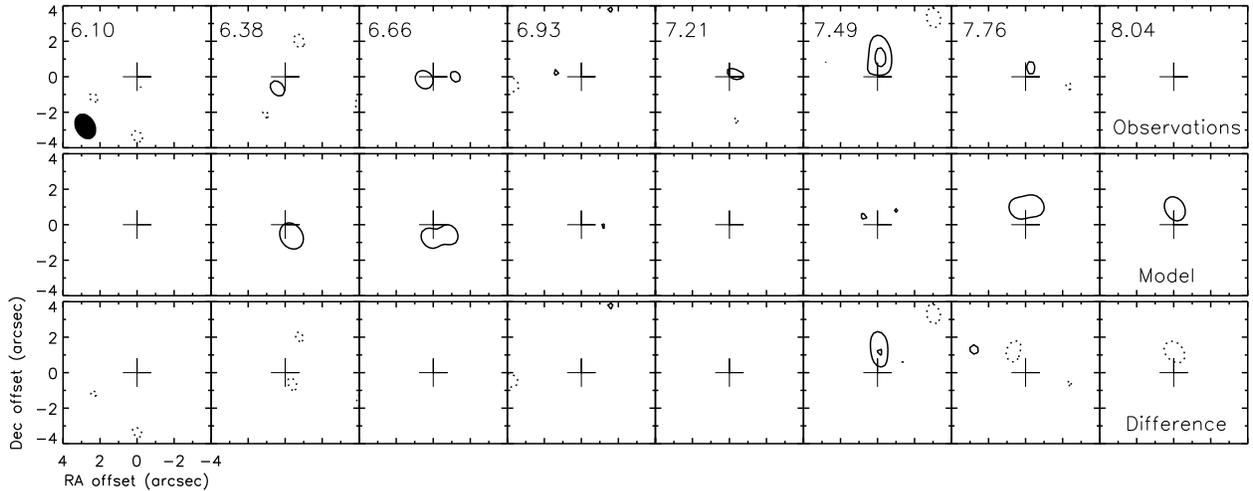}
\caption
      {-- {\it{Top panel}:} Channel maps of the observed C$^{18}$O $J$=2--1 emission at a range of velocities over which the line is detected. The lower left corner of this panel shows the size and position angle of the synthesized beam. -- {\it{Middle panel}:} C$^{18}$O $J$=2--1 channel maps from the best-fit disk model from Section 4.3.1. --  {\it{Lower panel}:} Channel maps showing the residual emission after subtraction of the best-fit model from the data. The contour levels are -1,1,2,3,4,5$\times$360~mJy~beam$^{-1}$ (2 sigma) in all panels.}
\label{18maps} 
\end{figure*}

Emission of $^{13}$CO and C$^{18}$O $J$=2--1 was detected from the HD169142 circumstellar disk. Figure~\ref{mom} shows the intensity weighted velocity maps with overlaid integrated intensity contours for both lines as well as the previously published $^{12}$CO $J$=2--1 line \citep{raman}. The intensity integrated over the velocity range from 5.6 to 8.4~km~s$^{-1}$, in which the line emission is fully contained, and over the central 4$\arcsec\times$4$\arcsec$ region is 12.1, 6.5 and 2.7~Jy~km~s$^{-1}$ for $^{12}$CO, $^{13}$CO and C$^{18}$O $J$=2--1 line, respectively. All three lines follow a similar velocity pattern, interpreted as a clear indication of Keplerian rotation around a 2~M$_{\odot}$ star of a disk seen at 13$^{\circ}$ inclination (Raman et al., 2006; this work).
Figure ~\ref{spec} shows the $^{12}$CO, $^{13}$CO and C$^{18}$O $J$=2--1 line spectra summed over a 4$\arcsec\times$4$\arcsec$ region centered at the HD169142 position. The profiles are relatively symmetric, double peaked and centered at 7.1$\pm$0.2~km~s$^{-1}$ and reflecting the underlying rotation pattern.
The $^{12}$CO and 1.3~mm results were presented by \citet{raman}. They report a 1.3~mm line flux of  169$\pm$5~mJy.
In Fig.~\ref{13maps} the emision stucture is shown over a range of velocities corresponding to the $^{13}$CO $J$=2--1 line (top panel). The emission extends to 2$\arcsec$ (270~AU) from the star at a 2$\sigma$ level and shows a Keplerian velocity pattern. Figure ~\ref{18maps} presents the channel maps of the C$^{18}$O $J$=2--1 line  (top panel) with a kinematic structure similar to that of the $^{13}$CO $J$=2--1 line seen in Fig. ~\ref{13maps}. The 2$\sigma$ level emission reaches 1$\farcs$5 (220~AU) from the star. At 7.4~km~s$^{-1}$ a localised emission peak of 0.94$\pm$0.35~Jy is seen 1$\farcs$0 north from the star.

\section{Discussion}

\citet{raman} show that the model that fits the SED of HD169142, can also be used to fit the structure and intensity of the resolved $^{12}$CO emission. They note that some weak residual emission is still present after subtracting the model from the observational data, that may correspond to real substructure within the disk. We analyse $^{13}$CO and C$^{18}$O data, which are expected to probe different depths in the disk due to their lower opacities.

\subsection{Adopted disk model}

\citet{dentHD} fit the spectral energy distribution (SED) of HD169142 using an accretion disk model from the \citet{dalessio} database with an accretion rate of 10$^{-8}$~M$_{\odot}$yr$^{-1}$, an outer radius of 300~AU, and a 30$^{\circ}$ inclination. The age of the central star in this model is 10~Myr. 
They adopt an A2 spectral type for the central star because it provides a slightly better fit than the A6 spectral type model does to the SED in the range of 2-200~$\mu$m dominated by the warm dust emission. However, there is no significant difference between the SEDs corresponding to these two models.
Beyond 44~AU from the star, the model is qualitatively similar to the model of the outer disk of HD169142 described in \citet{grady}. The detailed model structure, including plots of surface density distribution, temperature at different scale heights and optical thickness of the disk, is provided in the online database of accretion disk models \citep{dalessio}\footnote{www.astrosmo.unam.mx/~dalessio/}.

In order to match their resolved submillimeter observations of CO, \citet{raman} modify the model used in \citet{dentHD} to have a disk radius of 235~AU. Assuming the mass of the central star to be 2~M$_{\odot}$, they derive an inclination of 13$\pm$1$^{\circ}$ from face-on and use it in the model. These modifications are not expected to affect the quality of the SED fit since the SED of HD169142 alone does not constrain well the outer radius and inclination of the disk. They show that this SED model can be used to fit the structure and intensity of the resolved $^{12}$CO emission. However, the SED does not provide reliable constraints of the disk dust mass, which in the model is set to 2$\times$10$^{-4}$~M$_{\odot}$. This parameter depends only on the optically thin thermal emission from the disk in the millimeter wavelength region. 

The density in the model is a parameter that can be scaled by a factor of few, without significantly affecting the computed disk temperature structure. The resulting mass only affects the millimeter flux in the modeled SED, i.e., the 1.3~mm data point. In section~4.2 we re-analyse the 1.3~mm data from \citet{raman} using a radiative transfer code to calculate the emission from the model and better constrain the dust mass of the disk.

In the analysis of our $^{13}$CO and C$^{18}$O line emission (section~4.3) we use this improved disk model, which already encompasses a range of observational constraints. We assume that the gas follows the dust distribution prescribed by the model and that their temperatures are the same. The micro turbulent FWHM of the lines is set to 0.16~km~s$^{-1}$ throughout the disk. One conclusion that we can already draw from the disk model is the fraction of disk mass where freeze-out of CO is likely to be efficient, which is believed to happen at temperatures below 20~K. The mass weighted temperature in the model is 35~K, which indicates that the bulk of the disk material is at temperatures above 20~K. The temperature falls below 20~K only in the midplane region of the disk beyond 150~AU from the star and there the abundance of the gas phase CO and its isotopologues is expected to be heavily decreased due to freeze-out of these molecules onto dust grains. This region contains only 8\% of disk mass and therefore any mass estimate based on $^{13}$CO and C$^{18}$O emission may represent 92-100\% of the true disk mass. 

In all previously published work on HD169142, the disk gas mass was inferred indirectly from the derived disk dust mass, adopting a standard gas to dust ratio. In our analysis of the $^{13}$CO and C$^{18}$O emission we argue that it is reasonable to assume a standard $^{12}$CO abundance in the case of HD169142, because freeze-out affects less than 8\% of the mass in this disk. Therefore we investigate the constraints 
that can be obtained from our molecular line data on the disk gas mass. 

\subsection{Dust continuum emission}

From the observed 1.3~mm continuum flux of 169$\pm$5~mJy, \citet{raman} derive a dust mass of 2$\times$10$^{-4}$~M$_{\odot}$ assuming an emissivity $\kappa_{\nu}^{dust}$=2~cm$^2$g$\mathrm{^{-1}_{dust}}$ and a single temperature of 30~K. We reanalyse the 1.3~mm emission using the disk temperature and density structure from the model described in section 4.1, where we vary the dust density (i.e., the dust mass) to fit the observations. A large uncertainty on the mass determination from the 1.3~mm continuum flux is due to the dust opacity, which is not well determined in circumstellar disks. \citet{ossenkopf} study the effects of dust coagulation, and ice coverage of grains on their opacity in protostellar cores and suggest 1.3~mm opacity of 1~cm$^2$g$\mathrm{^{-1}_{dust}}$ for very dense (n$>$10$^{7}$~cm$^{-3}$) regions, and up to 5 times more if the grains are not covered in ice mantles. The latter is thus only valid for regions above water ice desorption temperature (80-100~K, \citet{fraser}). In their study of the dust opacity in circumstellar disks \citet{draine} find that astrosilicate and pirolysed cellulose at 600~$^{\circ}$C are materials representative of the dust properties which may be expected in circumstellar disks. They explore a large range of grain size distributions and find that these materials have mm-wavelength opacities which are close to the observational constraints on dust opacity from extinction studies of the diffuse interstellar medium presented in \citet{weingartner}.
The resulting opacity could be anywhere between 0.1 and 2~cm$^2$g$\mathrm{^{-1}_{dust}}$ at the wavelength of 1.3~mm, depending on the adopted grain size. 
We use this information to calculate the minimum amount of dust needed to produce the observed flux by adopting the opacity (and emissivity) of 2~cm$^2$g$\mathrm{^{-1}_{dust}}$. 
The SED modelling of the disk emission at shorter wavelengths is not affected by our assumptions of dust opacity at 1.3~mm. The dust thermal emission is dominated by the cold disk midplane at large radii, likely to contain settled and/or grown dust particles whose optical properties may differ from those of the small dust at disk surface, responsible for the optically thick near-infrared and infrared emission of the disk.

\begin{figure*}
\begin{center}
\parbox{15cm}{ \it }
  \begin{minipage}{1.0\textwidth}
        \includegraphics[width=8cm]{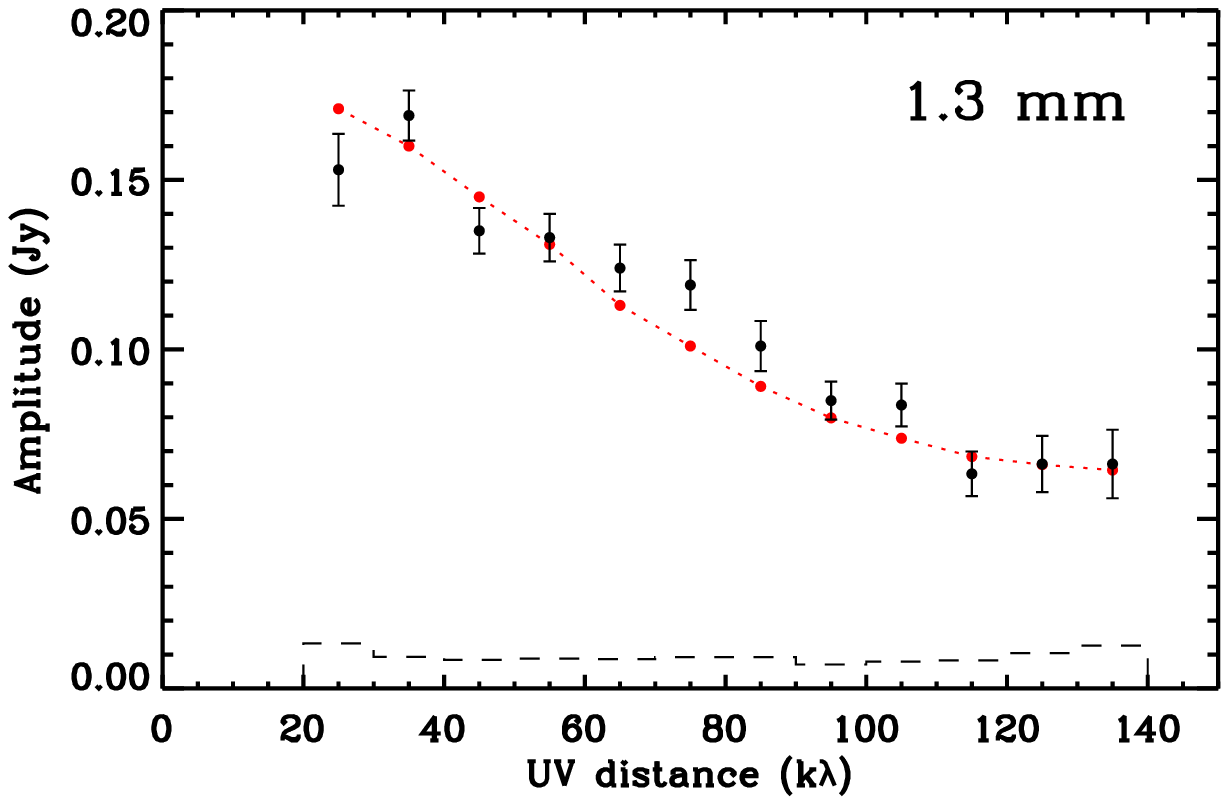}
    \includegraphics[width=8cm]{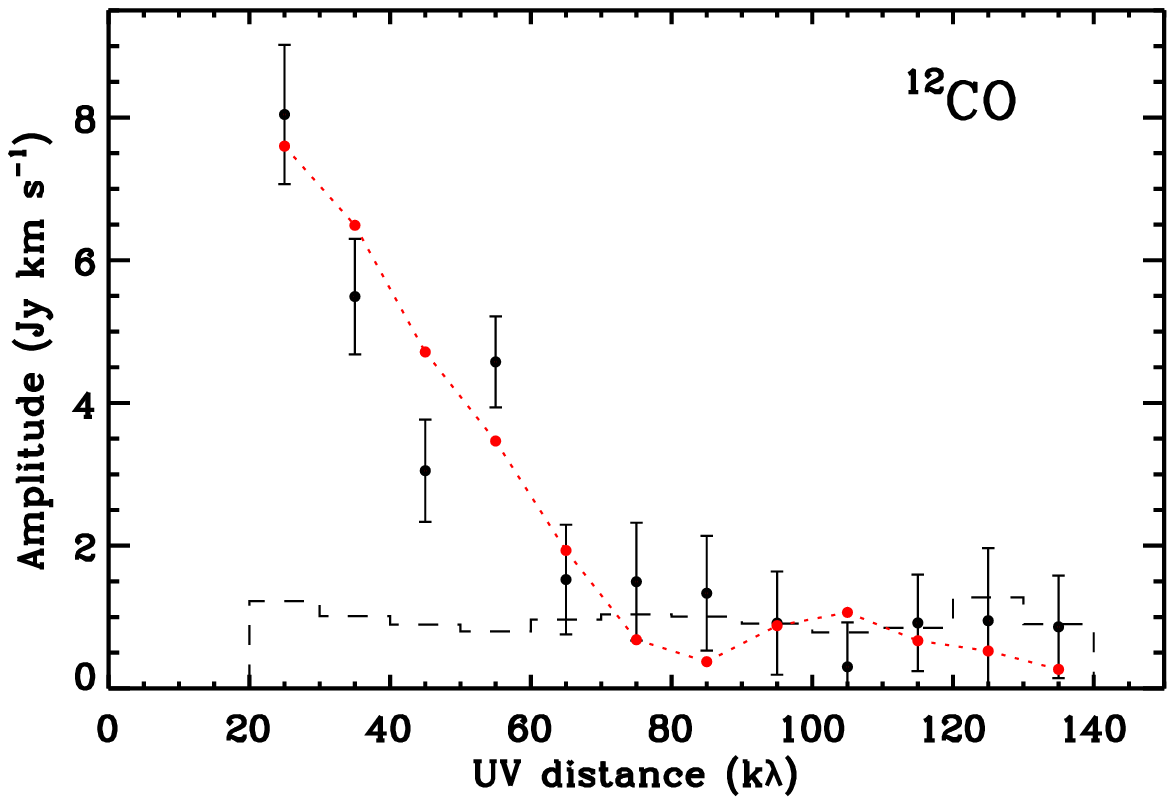}
  \end{minipage}
\parbox{15cm}{ \it }
  \begin{minipage}{1.0\textwidth}
        \includegraphics[width=8cm]{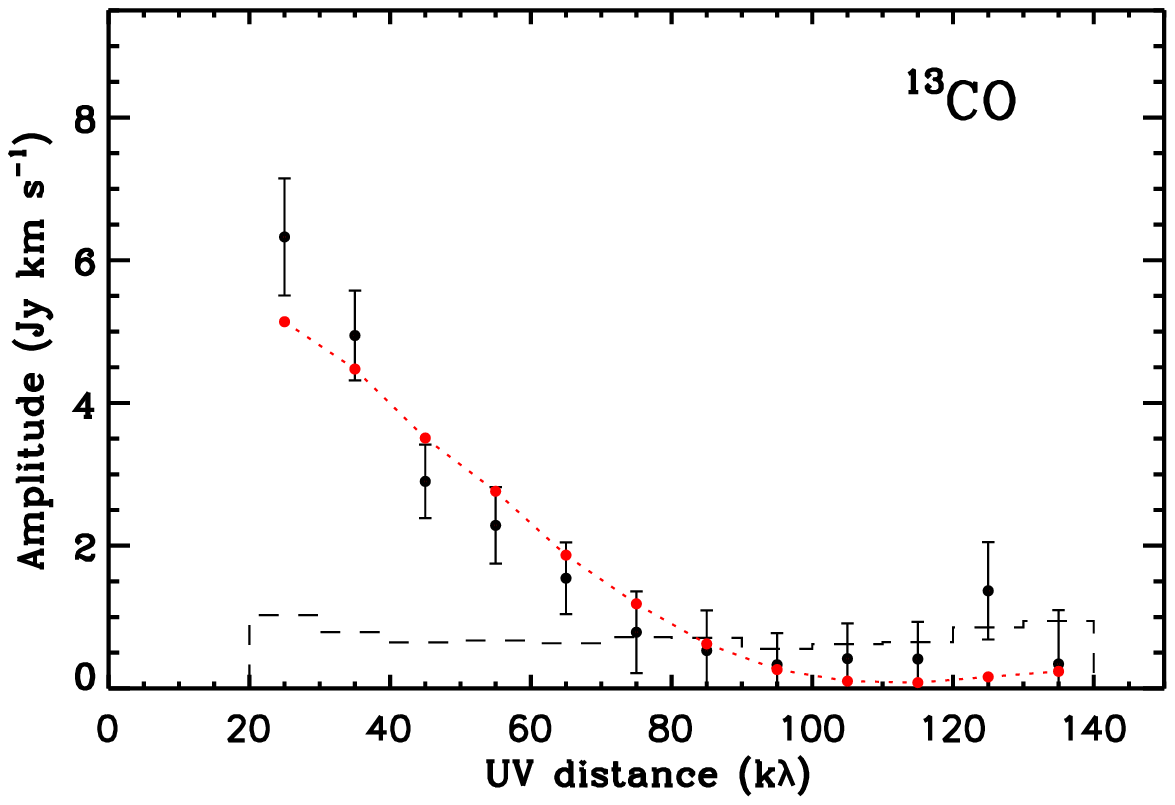}
    \includegraphics[width=8cm]{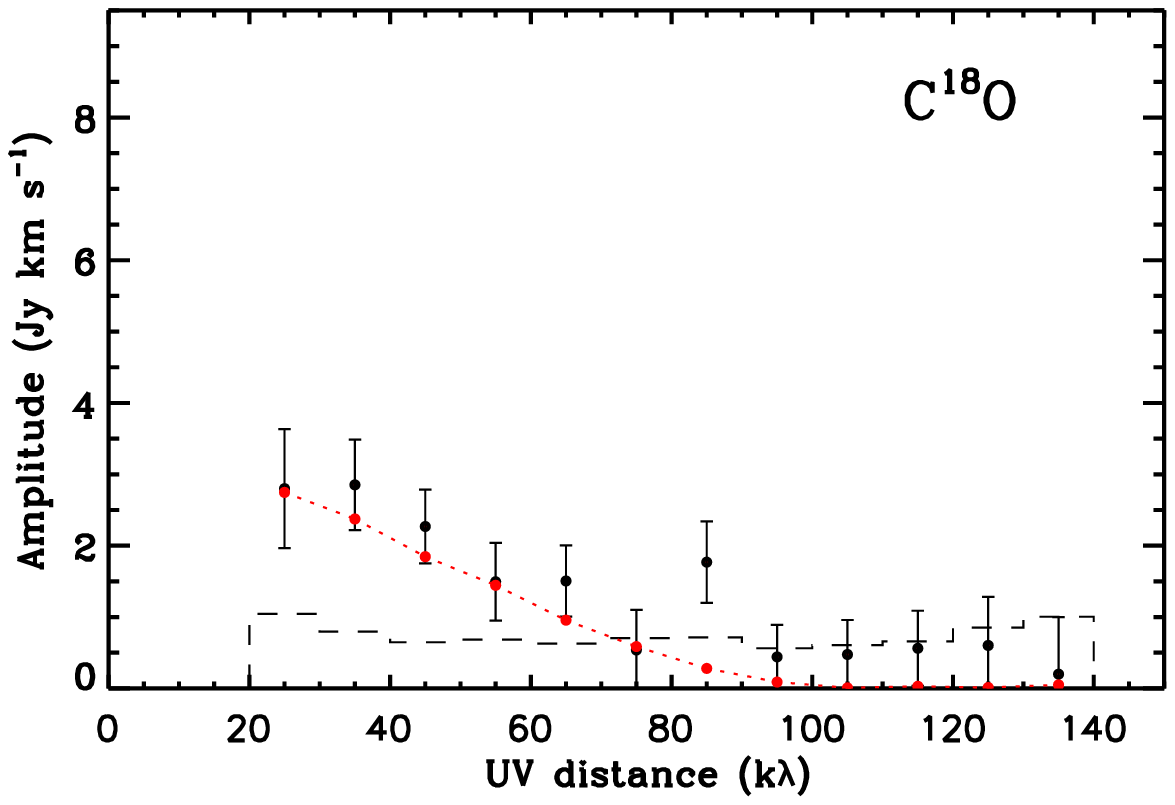}
  \end{minipage}
\caption{{\it{Top-left}:} Vector-averaged continuum flux as a function of projected baseline length (black dots). The error bars represent the variance within each annular average. The zero-signal expectation value is shown by the dashed histogram. The red dotted line shows the best-fit model for a dust emissivity of 2.0~cm$^2$g$\mathrm{^{-1}_{dust}}$ with the disk mass of 2.16$\times$10$^{-4}$~M$_{\odot}$. {\it{Top-right}:} Same for the $^{12}$CO emission from \citet{raman}, integrated over the width of $\approx$2.7~km~s$^{-1}$ (10 spectral channels) centered on each line. The red dotted line shows the model presented by \citet{raman}. {\it{Bottom-left and bottom-right}:} Same for our $^{13}$CO and C$^{18}$O J$=$2--1 data, respectively. The red dotted lines show the best-fit model from Section 4.3.1. }\label{uvamp}
\end{center}
\end{figure*}

We do the full modeling of the interferometric visibilities using the radiative transfer code of \citet*{hogerheijde} (RATRAN). We perform a $\chi^2$ minimisation to fit the 1.3~mm visibilities by varying the disk dust mass. The best match is shown in Fig.~\ref{uvamp}, top-left panel, resulting in an estimate of the dust mass of the disk, given by
\begin{equation}
\mathrm{M_{dust}= 2.16\times10^{-2}\frac{2~cm^2g{^{-1}_{dust}}}{\kappa_{1.3~mm}}\times M_{\odot}}.
\end{equation}
Assuming that the adopted dust opacity of 2~cm$^2$g$\mathrm{^{-1}_{dust}}$ is the maximum value, our estimate presents the lower limit on the dust mass in HD169142:
$\mathrm{M_{dust}\geq2.16\times10^{-4}~M_{\odot}}$.
A slightly higher maximum dust opacity of 7~cm$^2$g$\mathrm{^{-1}_{dust}}$ is obtained for pyrolised cellulose at 800~$^{\circ}$C but this value exceeds greatly the observational contraints from \citet{weingartner} and although not impossible, is unlikely in disks \citep{draine}. However we cannot exclude the possibility that other dust properties like cristallinity, porosity and shape, which are not well understood, may enhance dust submillimeter opacity beyond 2~cm$^2$g$\mathrm{^{-1}_{dust}}$ by a factor of a few.

\subsection{Molecular line emission}

Due to their low abundance relative to $^{12}$CO, the $^{13}$CO and C$^{18}$O molecules emit lines that are comparatively less optically thick. As a result of this, they saturate at different heights in the disk allowing us to use them as a probe of the physical conditions in different disk layers and investigate the disk vertical structure \citep{dartois}. Furthermore, the C$^{18}$O line emission is often optically thin in the outer disk region, where the bulk of disk mass is located, which makes it a valuable probe of the disk gas mass. 

The commonly employed route for the analysis of the CO isotopologue and dust continuum emission from circumstellar disks is to convert the continuum flux to disk mass adopting a certain dust opacity, usually 0.02~cm$^2$g$^{-1}_{\rm gas}$ \citep{beckwith}, under the assumption of the canonical gas to dust mass ratio of 100. In this way, the molecular line emission is fit for abundance, where any discrepancy with respect to the canonical abundances and isotopic ratios of observed molecules is explained by a depletion factor, indicating the fraction of the CO gas affected by the freeze-out onto dust grains or selective photodissociation. This approach is used in \citet{raman} where the analysis of the $^{12}$CO J$=$2--1 line emission and 1.3~mm continuum from HD169142 is done for a disk mass of 2$\times$10$^{-2}$~M$_{\odot}$ and $^{12}$CO was found to be depleted by a factor four, having an abundance of 2.5$\times$10$^{-5}$. This corresponds to  6$\times$10$^{-6}$~M$_{\odot}$ of $^{12}$CO molecules present in gas phase in the disk. 
In the present work, we argue that the $^{12}$CO abundance is not significantly affected by freeze-out or selective photodissociation and we chose to follow a different route, in which we derive a rough gas mass estimate from the molecular line observations taking into account all related uncertainties.

In order to interpret our $^{13}$CO and C$^{18}$O observations, we adopt the disk model described in detail in section 4.1. We set the dust mass in the model to 2.16$\times$10$^{-4}$~M$_{\odot}$; the value derived as the minimum dust mass for HD169142 in section 4.2. We assume that the molecular gas follows the dust temperature and density structure and we vary the $^{13}$CO and C$^{18}$O gas masses to best fit the observations.
We calculate the $^{13}$CO and C$^{18}$O $J$=2--1 line emission from the model using the RATRAN molecular excitation and radiative transfer code. The model visibilities are generated with the miriad task uvmodel, and the comparison to the observed visibilities is done directly. 
We use assumptions about the molecular abundances of the $^{12}$CO and isotopic ratios to make a conversion to the total gas mass (1.2$\times$M$_{\rm H_2}$). 
We explore a range of values of the gas to dust mass ratio f$_{g/d}$. 

\subsubsection{Disk structure probed by $^{13}$CO and C$^{18}$O}

\begin{figure*}
\centering
        \includegraphics[width=17cm]{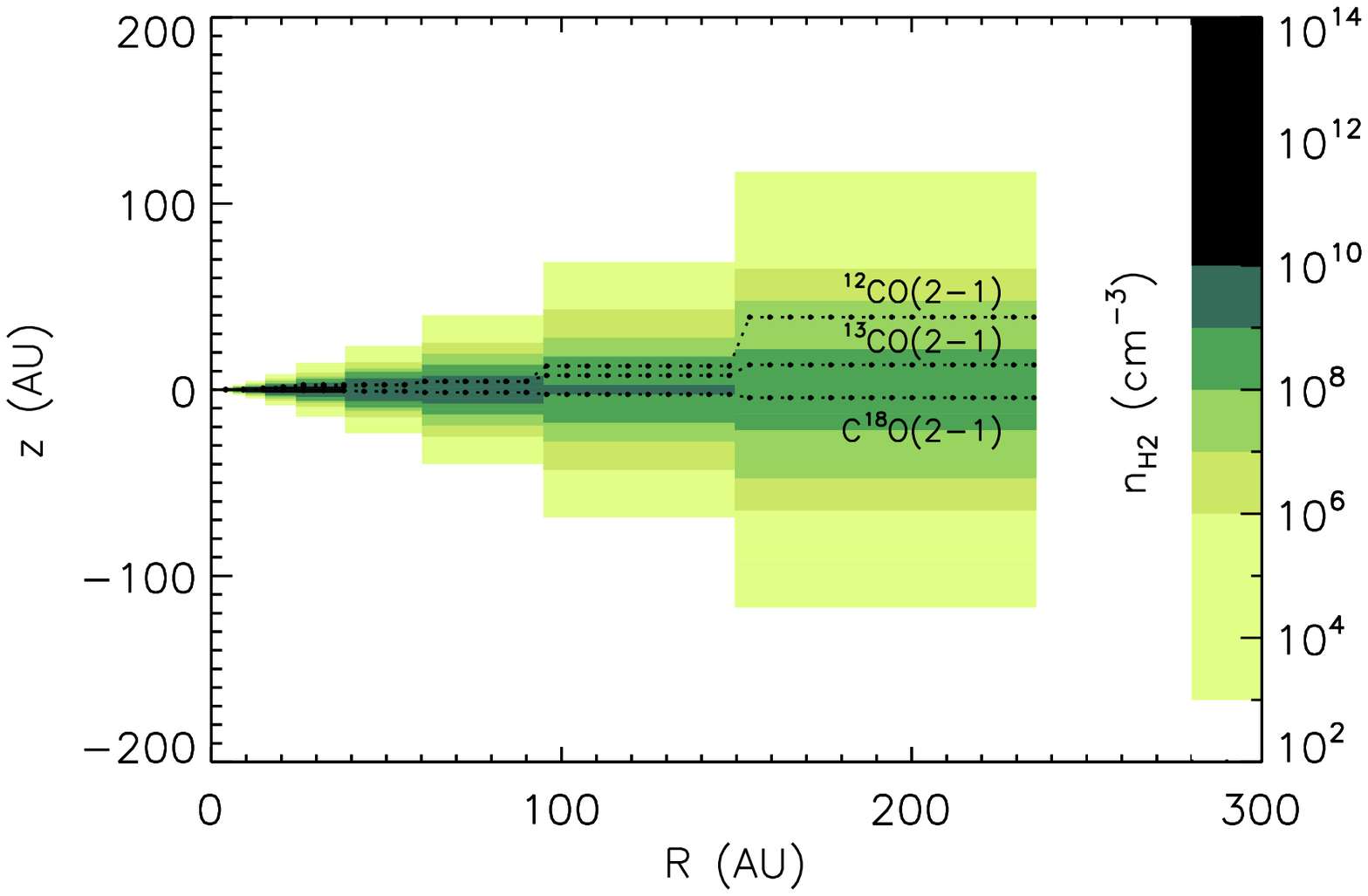}
\caption{Density structure of the best-fit disk model shown as colour scale. The disk is viewed by us from almost straight above. The three dotted lines show the heights above which 90\% of the integrated line emission of $^{12}$CO, $^{13}$CO and C$^{18}$O $J$=2--1 originates. For the C$^{18}$O emission this line lies below the disk midplane.}
\label{denstemp}
\end{figure*}

We find fits to the observed visibilities of $^{13}$CO and C$^{18}$O line for M$\mathrm{_{^{13}CO}}$=2.9$\times$10$^{-7}$~M$_{\odot}$ and M$\mathrm{_{C^{18}O}}$=4.6$\times$10$^{-8}$~M$_{\odot}$, shown in Fig.~\ref{uvamp}, bottom panels. For comparison, the visibilities of the $^{12}$CO J$=$2--1 line are also shown in Fig.~\ref{uvamp}, top-right panel, with the best fit obtained by \citet{raman} for M$\mathrm{_{^{12}CO}}$=6$\times$10$^{-6}$~M$_{\odot}$ contained in the gas phase.
In order to test our fit further, we invert and deconvolve the visibilities using the miriad reduction package and compare the observed (top panels) and modeled (middle panels) velocity channel maps in Figs.~\ref{13maps} and \ref{18maps}. The structure and intensity of the emission is found to be well matched. Figures ~\ref{13maps} and ~\ref{18maps} (lower panels) show the residual emission resulting from subtraction of the modeled from the observed data. For $^{13}$CO the residuals appear randomly distributed and do not exceed the noise level. Significant residuals are seen in C$^{18}$O at 7.4~km~s$^{-1}$ about 0$\farcs$5 north from the star. 
The C$^{18}$O data (Fig. ~\ref{18maps}) have a marginally significant localised excess $1\farcs0$ north of the star seen at 3~$\sigma$ level in the residual emission. This excess emission corresponds to the redshifted peak of the spectral line at v$_{\rm LSR}$=7.4~km~s$^{-1}$. It is marginally resolved at $1\farcs5\times2\farcs5$ and contains about 20\% of C$^{18}$O line flux. 
Similar, but not spatially coincident residual emission is also seen in $^{13}$CO (Figure ~\ref{13maps}, lower panel) near the noise levels, but not in $^{12}$CO. A cause of this feature may lie in disk midplane asymmetry. Any connection with the reported companion at 9$\farcs$3 is unlikely, but a hypothesized body within the inner disk is an interesting prospect \citep{grady}. Nevertheless, an instrumental or calibration artefact may have affected the 7.4~km~s$^{-1}$ channel resulting in a slight increase of flux in that channel in the lower sideband, only noticeable in the weak C$^{18}$O line.

In this section we will use our $^{13}$CO and C$^{18}$O results to make an estimate of the total gas mass M$_{gas}$ in the disk, related to the $^{12}$CO gas mass M$_{^{12}CO}$ through
\begin{equation}
\mathrm{M_{gas}=1.2\times\frac{m_{H_2}}{m_{^{12}CO}}\times\frac{1}{[^{12}CO]}\times{M_{^{12}CO}}}.
\end{equation}

where $\mathrm{m_{H_2}}$ and $\mathrm{m_{^{12}CO}}$ are the masses of the H$_2$ and $^{12}$CO molecule, and [$^{12}$CO] is the abundance of $^{12}$CO with respect to H$_2$. The factor of 1.2 accounts for one fifth of the gas contained in helium.
We can use the derived masses of the $^{13}$CO and C$^{18}$O gas to calculate the mass of the $^{12}$CO gas. We adopt the isotopic ratios of $^{12}$C/$^{13}$C=77$\pm$7 and $^{16}$O/$^{18}$O=560$\pm$25 from \citet{wilson}.
\begin{equation}
\mathrm{M_{^{12}CO}=\frac{m_{^{12}CO}}{m_{C^{18}O}}\times\frac{[^{16}O]}{[^{18}O]}\times{M_{C^{18}O}}=(2.4\pm1.0)\times10^{-5}M_{\odot}}
\end{equation}
\begin{equation}
\mathrm{M_{^{12}CO}=\frac{m_{^{12}CO}}{m_{^{13}CO}}\times\frac{[^{12}C]}{[^{13}C]}\times{M_{^{13}CO}}=(2.2\pm0.6)\times10^{-5}M_{\odot}}
\end{equation}
The values obtained here are given with errors which are dominated by the rms of our data and also include the errors on isotopic ratios. If we take into account that 8\% of the $^{12}$CO and its isotopologues in HD169142 may be depleted onto dust grains and therefore not contributing to the emission observed, then the full $^{12}$CO mass range is (1.4-3.7)$\times$10$^{-5}$~M$_{\odot}$. The fact that both datasets roughly agree on the $^{12}$CO mass provides an additional argument for the validity of the model. However, the fit to the $^{12}$CO J$=$2--1 data done in \citet{raman} corresponds to a mass of gas phase $^{12}$CO of 6$\times$10$^{-6}$~M$_{\odot}$ (mentioned in section 4.3), two to six times lower than the range derived here. 
The $^{12}$CO emission is optically thick and therefore it does not trace the entire disk but originates from a layer much higher in the disk than either the $^{13}$CO or C$^{18}$O lines \citep{dartois}. The discrepancy between the $^{12}$CO result and that of the isotopologues indicates that the adopted disk model may not provide the best description in these high layers. We will discuss this in more detail in section 4.5 where the effects of temperature and turbulence on the line profile are discussed.

Our model calculations allow us to identify which regions of the disk contribute most to the observed lines. We follow the line of sight integration step by step from the observer into the disk and find the location where 90\% of the integrated intensity is reached for a face on orientation of the disk. We show these locations in Fig.~\ref{denstemp} in a cross-section of the disk showing the underlying density structure. We choose to use the intensity integrated over a range of wavelengths in which the line is emitted in order to better represent the region corresponding to the physical quantity that is observed, i.e., the line flux. The use of surfaces with equal opacity (often used $\tau$=3) may be misleading, because the opacity at the line centre is not necessarily consistent with the opacity in the line wings, and the amount of flux still produced in the line wings beyond the $\tau$=3 surface may contribute significantly to the line flux in some species more than in the others. The calculation was done for $^{12}$CO, $^{13}$CO and C$^{18}$O J$=$2--1 line. The discontinuous shape of the surface above which 90\% of the integrated intensity in these lines is emitted is due to the discrete cells in the grid of our model that are used to sample the density and temperature structure. As expected, the less optically thick species trace regions deeper in the disk with C$^{18}$O tracing the disk midplane while $^{12}$CO traces the disk surface layer. 
We analyse the physical properties of the emitting regions in the outer disk, beyond 100~AU from the star (scales resolved by our SMA observations). For the two outermost radial positions in our model grid, we calculate the column of $^{12}$CO gas contained in the region where 90\% of the emission is coming from, which is located above the vertical position shown in Fig.~\ref{denstemp}, for each molecular line. These values are shown in Table~\ref{table}\footnote{The calculations for $^{12}$CO J$=$2--1 are done for the model presented in \citet{raman}}. Also, the average temperature of these gas columns is calculated, taking into account the temperature weighted by the column density (i.e., mass) for each cell located above the vertical positions from Fig.~\ref{denstemp}. 
The C$^{18}$O J$=$2--1 line generally traces three times as much CO gas as traced by the $^{13}$CO J$=$2--1 line in the outer disk while the $^{12}$CO J$=$2--1 line traces only a small fraction of the gas, a few percent of the column traced by C$^{18}$O. It is clearly seen that the C$^{18}$O traces deep into the disk, to the height of several AU at the far side of the midplane, showing that it is very sensitive to the total gas mass. The 90\% of the C$^{18}$O J$=$2--1 line emission traces around 60\% of the total gas column. 

\begin{table*}
 \caption{The table shows the column of $^{12}$CO traced by 90\% of the emission of each line, and the corresponding average temperature, at two radial intervals beyond 100~AU. These values are representative of the regions shown in Fig.~\ref{denstemp}. The column densities of $^{12}$CO contained above the vertical locations z from Fig.~\ref{denstemp} are shown. The average temperature values shown are column-weighted (i.e. mass-weighted) values over all cells which are located above the reported height z. }
\label{table}
\centering
\begin{tabular}{c | c c | c c c c | c c}
 \hline\hline
Molecular & z(AU) & z(AU) & N$_{\rm CO}$(cm$^{-2}$) & Fraction Of & N$_{\rm CO}$(cm$^{-2}$) & Fraction Of & T$_{\rm ave}$ (K) & T$_{\rm ave}$ (K)\\
Line & 100-149~AU & 149-235~AU & 100-149~AU & Total N$_{\rm CO}$  & 149-235~AU & Total N$_{\rm CO}$ & 100-149~AU & 149-235~AU\\
\hline
$^{12}$CO J$=$2--1 & 12.7 & 39.0 & 6.5$\times$10$^{17}$ & 7\% & 6.6$\times$10$^{16}$ & 1\% &33.3 &38.6\\
$^{13}$CO J$=$2--1 & 7.6 & 13.0 & 4.2$\times$10$^{18}$ & 17\% & 2.6$\times$10$^{18}$ & 18\% & 29.1 & 25.6\\
C$^{18}$O J$=$2--1 & -2.5 & -4.3 & 1.5$\times$10$^{19}$ & 62\% & 8.9$\times$10$^{18}$ & 62\% & 25.7 & 21.8\\
\hline
\end{tabular}
\end{table*}

In section 4.1 we already concluded from the disk model that the CO freeze-out could only be efficient in the cold outer disk midplane region which contains 8\% of the total disk mass. This very small fraction is due to the luminosity of the Herbig Ae star HD169142 which is higher than the typical luminosity of T Tauri stars. In the latter, the circumstellar disks are colder and the CO freeze-out presents a major obstacle to determine gas mass based on observations of CO or its isotopologues. As a caveat, we note that \citet{pietu3} find somewhat lower dust temperature in the disk around the Herbig Ae star MWC 480 (10-25~K at 100~AU). Since we have no indications to the contrary, we adopt the dust temperature that SED disk modeling suggests.
If observed with a sufficiently high resolution and sensitivity, disks where CO freeze-out is efficient are expected to have smaller radii derived from CO isotopologues compared to the outer radius derived from the $^{12}$CO and from the dust emission. Although our observations are limited by the beam size (1$\farcs$4$\approx$200~AU) we have tested a range of different outer radii and found that, based on our molecular line data, the outer radius should not be very different from 200~AU for either of the observed lines. 
We also do not find any significant discrepancy between the molecular line data and dust continuum that could indicate different outer radii for gas and dust. 

Another factor that can influence the $^{12}$CO abundance is photodissociation. The effect of photodissociation via the stellar UV radiation on CO abundance in circumstellar disks is studied in \citet{vanzadelhoff} and \citet{vanzadelhoffpaper}. They find that the CO abundance decrease due to photodissociation is significant only in the upper disk layers and that the CO column density is not much affected by it. 
\citet{jonkheid} show that the stellar UV field is only able to heat 10\% of the gas mass in the disk to considerably high temperatures, in the surface disk layer.
The effect of photodissociation on the CO content of HD169142 is expected to be small, due to the significant dust mass of the disk. Since we adopt a maximum emissivity and therefore minimal dust mass, the conclusion that photoionization can be neglected is robust.
In addition, the C$^{18}$O emission that we use to derive the gas mass, arises predominantly from the outer disk midplane region that contains the bulk of the disk mass and does not receive any direct UV illumination from the central star. This minimises further the effects of photodissociation on our results. Therefore, we find it appropriate to use the $^{12}$CO mass derived from our best fit to the C$^{18}$O and $^{13}$CO data to make an estimate of the total gas mass. 

The abundance of $^{12}$CO is a key factor in this conversion and the values found in the literature for the interstellar medium range from 9.5$\times$10$^{-4}$ \citep{frerking} to 2.0$\times$10$^{-4}$ \citep{lacy}. 
If we adopt the $^{12}$CO abundance of 10$^{-4}$ with respect to H$_2$, we obtain $\mathrm{M_{\rm gas}=1.2\times M_{\rm H_2}=(2.0^{+1.0}_{-0.8})\times10^{-2}M_{\odot}.}$
The error estimate includes the error due to the rms of our C$^{18}$O data (41\%) and to the isotopic ratio of [$^{16}$O]/[$^{18}$O] (4\%). The upper margin is extended to include the 8\% uncertainty in the gas mass estimate due to freeze-out. The factor 1.2 takes into account the gas mass contained in helium, the second most important gas mass constituent after molecular hydrogen. Adopting a $^{12}$CO abundance of 2.0$\times$10$^{-4}$ with respect to H$_2$ results in M$_{\rm gas}$ as low as 6$\times$10$^{-3}$~M$_{\odot}$. Therefore, due to the uncertainty largely dominated by the $^{12}$CO abundance, we can only conclude that the gas mass of the disk around HD169142 falls in the range of (0.6-3.0)$\times$10$^{-2}$~M$_{\odot}$.

\subsection{Gas to dust ratio}

\begin{figure}
 \includegraphics[width=8cm,angle=0]{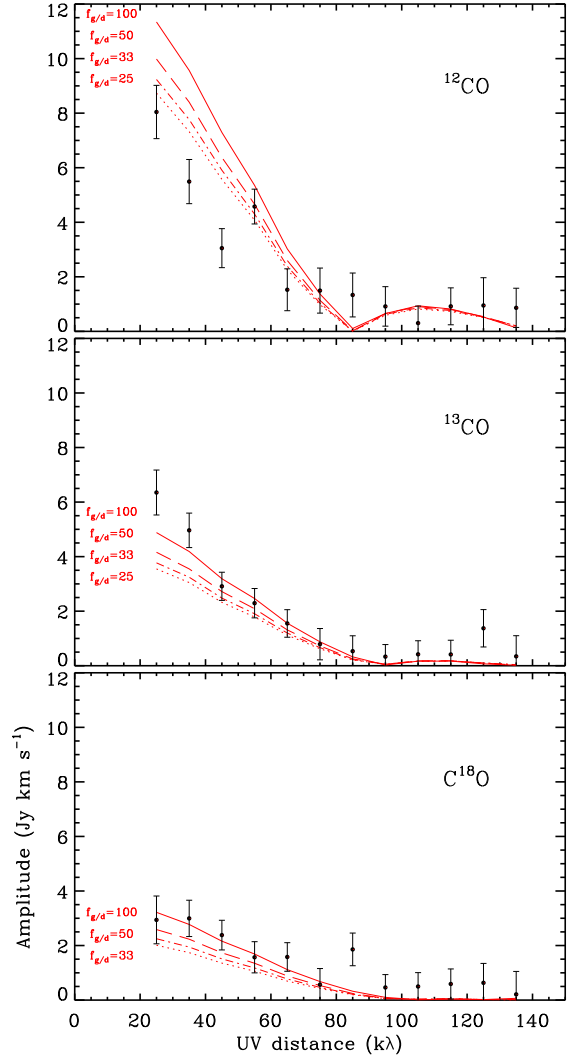}
\caption{Comparison of the observed vector-averaged $^{12}$CO ({\it{top panel}}), $^{13}$CO ({\it{middle panel}}) and C$^{18}$O ({\it{lower panel}}) line fluxes (black) to four models with different gas to dust ratios (red lines). The error bars represent the variance within each annular average. The models assume the dust mass of 2.16$\times$10$^{-4}$~M$_{\odot}$ and gas to dust mass ratio f$_{\rm g/d}$=100,50,33,25 marked respectively by the full, dashed, dashed-dotted and dotted red lines in each panel. The line fluxes are integrated over the width of $\approx$2.7~km~s$^{-1}$ (10 spectral channels) centered on each line. }\label{uvamp2}
\end{figure}

In order to investigate the gas to dust ratio, we ran several models in which we assume M$\mathrm{_{\rm dust}}$=2.16$\times$10$^{-4}$~M$_{\odot}$, [CO]=10$^{-4}$, $^{12}$C/$^{13}$C=77, $^{16}$O/$^{18}$O=560 and gas to dust mass ratios of 100, 50, 33 and 25. From the comparison of the synthetic visibilities of these models with the observations (Fig.~\ref{uvamp2}, middle and lower panel) we find that a good fit for $^{13}$CO and C$^{18}$O data is obtained for f$_{\rm g/d}$=100. Under the assumptions used above, lower values of f$_{\rm g/d}$ fail to provide a fit to the data. 
The $^{12}$CO data (Fig.~\ref{uvamp2}, top panel) shows a discrepancy, already discussed in section 4.3.1, with respect to the $^{13}$CO and C$^{18}$O data in the sense that the model we use overestimates the $^{12}$CO emission. This suggests that the model assumptions like the temperature, micro turbulence and $^{12}$CO abundance used for the upper disk layers where the $^{12}$CO emission arises from should be reconsidered. 
From the gas and dust masses derived in sections 4.2 and 4.3 we see that a gas to dust mass ratio of 100:1 only holds for special circumstances considered here: a maximum dust emissivity of 2~cm$^2$g$\mathrm{^{-1}_{\rm dust}}$ at 1.3~mm and a $^{12}$CO abundance of 10$^{-4}$ with respect to H$_2$, or lower. For equally likely values of 1~cm$^2$g$\mathrm{^{-1}_{\rm dust}}$ for the dust emissivity and 2$\times$10$^{-4}$ for the $^{12}$CO abundance, a gas to dust mass ratio of 25 would be inferred, suggesting that the disk would have lost significant amounts of gas with respect to dust. 
In fact, gas loss is expected in circumstellar disks due to photoevaporation of gas by the stellar radiation. The typical timescale for this process of a few Myr \citep{hollenbach,font} is similar to the estimated age of HD169142 of 6$^{+6}_{-3}$~Myr \citep{grady}. 

\subsection{Micro turbulence}

As mentioned in the previous section, the overestimate of the $^{12}$CO emission by the model which fits well the $^{13}$CO and C$^{18}$O $J$=2--1 line emission, may be due to model assumptions used for the upper disk layers, e.g., temperature, $^{12}$CO relative abundance, turbulence.

In section 4.1 we mention that a disk model with a less luminous A6 spectral type star also reproduces the SED of HD169142 well \citep{dentHD}. However, the temperature in that model is less than 10\% lower than the temperature used in our model. This is insufficient to provide a significant decrease in the $^{12}$CO line flux. 

A fit to all observed lines simultaneously by varying the isotopic ratios, yields an unusualy low value of both the $^{12}$C/$^{13}$C and the $^{16}$O/$^{18}$O isotopic ratio: 21 and 171, respectivelly. These values are far from the observationally derived ratios in the interstellar medium and this scenario would be highly unlikely. An over-abundance of $^{13}$CO was measured at large radial distances from the star (several hundreds of AU) in the disks DM Tau, MWC 480 and LkCa 15 \citep{pietu1}
In general, disks are often found to have $^{13}$CO and C$^{18}$O abundances depleted by large factors, mainly in disks around T Tauri stars \citep{dutrey1,dutrey2}.

\begin{figure}
\includegraphics[width=8cm,angle=0]{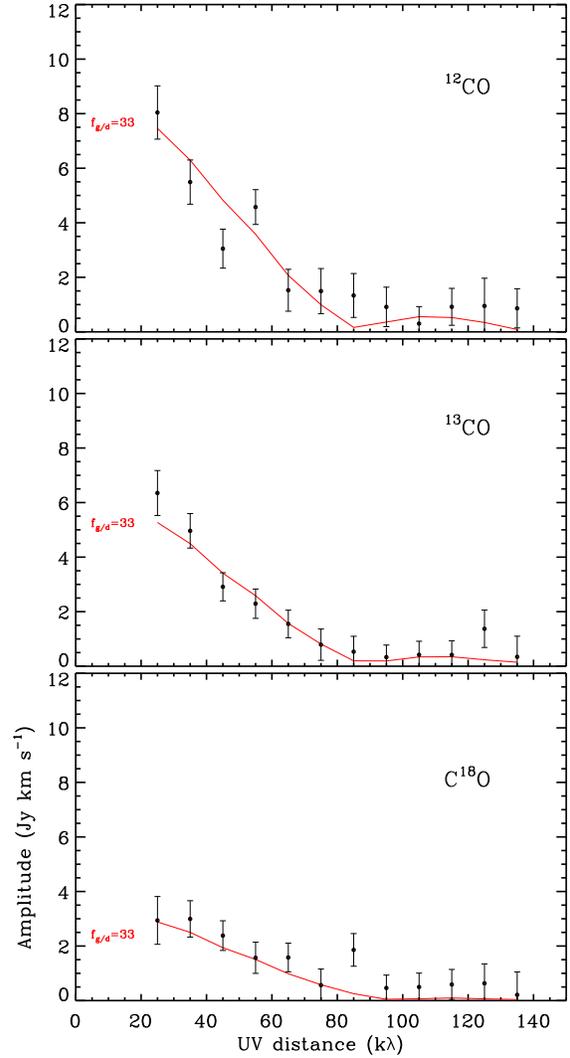}
\caption{Comparison of the observed vector-averaged $^{12}$CO ({\it{top panel}}), $^{13}$CO ({\it{middle panel}}) and C$^{18}$O ({\it{lower panel}}) line fluxes (black) to a model which assumes the total disk mass of 7$\times$10$^{-3}$~M$_{\odot}$, gas to dust mass ratio f$_{\rm g/d}$=33, no microturbulent line broadening in the upper disk layers from which the $^{12}$CO emission originates (only thermal broadening), and a microturbulent broadening of 0.33~km~s$^{-1}$ FWHM in the layers from which the $^{13}$CO and C$^{18}$O emission arises.}\label{turb}
\end{figure}

In the above analysis, the turbulent line FWHM was set to 0.16~km~s$^{-1}$ throughout the disk. In a more realistic scenario, the turbulence in the cold and dense midplane differs from the turbulence in the warmer and less dense surface layers. We investigate the effect of micro turbulence on the modeled line emission and find that a simultaneous match is possible in our model, as shown in Fig.~\ref{turb}. This fit is obtained for a gas to dust ratio of 33, i.e., a disk gas mass of 7$\times$10$^{-3}$~M$_{\odot}$, if a higher value of turbulence is assumed for the disk midplane (turbulent FWHM of 0.33~km~s$^{-1}$) and no turbulence for the upper disk layers traced by the $^{12}$CO emission (0.00~km~s$^{-1}$, only thermal broadening). For the $^{12}$CO J$=$2--1 line, that is fully optically thick at all radii in our model, the turbulent line broadening is degenerate with temperature and density in its effect on the line flux and profile. This is because the turbulence additionally broadens the thermally broadened line profile and in this way the optical thickness decreases allowing for more gas column to be traced. On the other hand, by decreasing the turbulence, at a given temperature, the same $^{12}$CO data can be fit with comparably more M$_{\rm CO}$. The effect of the increase in turbulence on the $^{13}$CO and C$^{18}$O lines is present but less pronounced, due to their lower optical thickness. In particular, the C$^{18}$O $J$=2--1 line emission is only marginally optically thick. The effect of turbulence on the modeled C$^{18}$O $J$=2--1 line emission is below the rms of our observations. It is not possible to obtain simultaneous fits to all data if the gas to dust ratio is higher than 33 (disk mass higher than 7$\times$10$^{-3}$~M$_{\odot}$). A lower gas to dust ratio (disk mass lower than 7$\times$10$^{-3}$~M$_{\odot}$) would require the turbulent FWHM of the lines to be larger than 0.33~km~s$^{-1}$ in disk midplane.

\section{Summary}

The circumstellar disk around the Herbig Ae star HD169142 was imaged in $^{13}$CO and C$^{18}$O $J$=2--1 line emission at 1$\farcs$4 resolution. The disk is resolved and the emission extends to 200-300~AU from the star at 2$\sigma$ levels. We adopt an accretion disk model which matches the SED of HD169142 and we use a molecular excitation and radiative transfer code to analyse our data. We revisit existing 1.3~mm and $^{12}$CO data and interpret it in the view of our new results. The observed kinematic pattern and line profiles are consistent with a disk seen at 13$^{\circ}$ inclination from face on, in Keplerian rotation around a 2~M$_{\odot}$ star. The structure of the emission suggests a disk radius of roughly 200~AU for both gas and dust. 
\begin{itemize}

\item We calculate the minimum disk dust mass of M$\mathrm{_{\rm dust}}$=2.16$\times$10$^{-4}$~M$_{\odot}$ through the fit to the 1.3~mm data using the maximum dust opacity theoretically predicted in circumstellar disks.

\item The adopted disk model reproduces our molecular line observations with M$\mathrm{_{C^{18}O}}$=4.6$\times$10$^{-8}$~M$_{\odot}$ and M$\mathrm{_{^{13}CO}}$=2.9$\times$10$^{-7}$~M$_{\odot}$. The derived amount of C$^{18}$O implies a M$\mathrm{_{\rm CO}}$=(2.4$\pm$1.0)$\times$10$^{-5}$~M$_{\odot}$. The $^{13}$CO amount is also in agreement with this value. 

\item We identify the regions of the disk that contribute most to the emission of $^{12}$CO, $^{13}$CO and C$^{18}$O J$=$2--1 line and find, as expected due to different optical depth of these lines, that they are physically distinct. In particular, the C$^{18}$O J$=$2--1 line is marginally optically thin throughout the disk and traces the bulk of the mass contained in the disk midplane. 

\item We find that the effect of freeze-out in the model is relatively small, allowing us to make a gas mass estimate based on our observations. For plausible CO abundances of (1-2)$\times$10$^{-4}$, we can only derive a range of M$_{\rm gas}$=(0.6-3.0)$\times$10$^{-2}$~M$_{\odot}$.

\item The comparison of the derived gas and dust mass shows that the gas to dust mass ratio of 100 is only possible under assumption of a high dust opacity of 2~cm$^{2}$g$^{-1}_{dust}$ and low $^{12}$CO abundance of 10$^{-4}$ with respect to H$_2$. Otherwise, this ratio may be as low as 25.

\item Existing observations of the optically thick $^{12}$CO line can be fit simultaneously to our $^{13}$CO and C$^{18}$O data by the model adopted here, only if the micro turbulence in the disk midplane is higher than the micro turbulence in the upper disk layers which dominate the $^{12}$CO emission. 
\end{itemize}

The discrepancy between the CO gas mass derived here and that derived based on the $^{12}$CO line emission likely implies that the micro turbulence is much lower than 0.1~km~s$^{-1}$ in the upper disk layers where the $^{12}$CO emission originates. 
Better constraints of the $^{12}$CO abundance in disks are needed in order to enable us to measure gas mass more precisely in disks. Better knowledge of dust properties such as dust size distribution and composition is essential for a better estimate of dust opacity in disks. However, these properties vary from source to source and are not necessarily linked with other parameters, age or mass for example and the studies need to focus on individual sources. Since the disks around Herbig Ae stars have the advantage that the gas mass estimate from C$^{18}$O line emission is more reliable, it would be useful to investigate dust properties, i.e., emissivity in these sources in more detail to be able to have a reliable estimate of dust mass as well. This would, for the first time, open a real opportunity to study the gas to dust mass ratio, the unknown property on which the theories of disk evolution and planet formation depend strongly.

\begin{acknowledgements}
The research of O.~P. is supported through Marie Curie FP6 programme of the European Union. The research of O.~P. and M.~R.~H. is supported through a VIDI grant from the Netherlands Organisation for Scientific Research. We would like to thank E.~F. van Dishoeck for useful discussions.
\end{acknowledgements}

\bibliographystyle{aa}
\bibliography{refs}
\end{document}